\documentclass[aps,prl,reprint]{revtex4-1}%
\usepackage{epsfig,pslatex,latexsym,times,amssymb,amsmath,graphicx}
\usepackage{bm}
\usepackage{amsmath}
\usepackage{amsfonts}
\usepackage{amssymb}
\usepackage{mathrsfs}
\usepackage{color}
\usepackage{graphicx}%
\providecommand{\U}[1]{\protect\rule{.1in}{.1in}}
\begin{document}
\author{N. J. Harmon}
\email{nicholas-harmon@uiowa.edu} 
\author{M. E. Flatt\'e}
\affiliation{Department of Physics and Astronomy and Optical Science and Technology Center, University of Iowa, Iowa City, Iowa
52242, USA}
\date{\today}
\title{Spin-flip induced magnetoresistance in positionally disordered organic solids}
\begin{abstract}
A model for magnetoresistance in positionally disordered organic materials is presented and solved using percolation theory. The model describes the effects of spin flips on hopping transport by considering the effect of spin dynamics on an effective density of hopping sites. Faster spin-flip transitions open up `spin-blocked' pathways to become viable conduction channels and hence produces magnetoresistance. The magnetoresistance can be found analytically in several regimes, including when the spin-flip time is slower than the hopping time. The ratio of hopping time to the hyperfine precession time is a crucial quantity in determining the shape of magnetoresistance curves. Studies of magnetoresistance in known systems with controllable positional disorder would provide a stringent test of this model. 
\end{abstract}\maketitle

Spintronics \cite{Ziese2001} in organic materials has generated considerable interest in recent years \cite{Naber2007}, due to the long spin lifetimes of organic semiconductors as well as the flexibility, low cost and chemical tunability of organic devices \cite{Hauff2006}.  
Spin transport properties are intimately connected to the electrical transport properties \cite{Flatte2000b}, so although spin transport through inorganic semiconductors has been extensively explored \cite{Awschalom2002, Awschalom2007}, novel features should be expected in organics due to their very different electronic transport properties.
The understanding of spin transport  in organics has been challenged by the discovery of magnetic field effects on properties such as conductivity and electroluminescence \cite{Kalinowski2003,Francis2004,Prigodin2006,Desai2007,Hu2007,Bloom2007,Bobbert2007,Bergeson2008,Wagemans2010}; and characterized  by magnetoresistances of 10-20\% in magnetic fields as small as 10~mT. 
Several new models of organic magnetoresistance (OMAR) have been proposed, many of which involve spin-dependent processes emanating from hyperfine interactions. These models can be broadly categorized by their reliance on bipolaron \cite{Bobbert2007} or electron-hole pair \cite{Prigodin2006, Desai2007} formation. Their main points of difference are that bipolaron models consider the relative spins of like charge carriers and electron-hole models consider the relative spins of electrons and holes on different sites in the formation probability for a two-carrier entity. However, no model of OMAR has explicitly taken into account how the presence of spin-blocked sites affects the theoretical description of hopping transport for a single carrier using percolation theory \cite{Shklovskii1984,Baranovski2006}.
 
This Letter provides a description of magnetoresistance based on percolative hopping transport for positionally disordered organic semiconductors. The model proposed here maps the complex phenomena of spin-dependent hopping onto a simple problem of $r$-percolation with an effective density of hopping-accessible sites that depends on the magnetic field through spin relaxation.
We focus on unipolar charge transport since several analytic results can be readily obtained; extension to bipolar transport can be done with similar techniques.
Our percolation based model allows us to explain the width and saturation of measured magnetoresistance curves as well as make predictions of magnetoresistance in systems with low site concentration and high temperatures. Finally, we propose experiments that could test our theory and thereby shed light on the operative mechanisms leading to OMAR. 

\emph{Model} - We model the spatially disordered organic system as a network of random resistors in the spirit of Miller and Abrahams \cite{Miller1960}. The resistance between two sites, $i$ and $j$, is given by $R_{ij} = R_0 e^{2 r_{ij}/a}$ where $r_{ij}$ is the their separation and $a$ is the localization length of a carrier at a site which we assume to be constant throughout the system. The bulk resistance in such a random resistor network is solved by percolation theory \cite{Ambegaokar1971, Pollak1972, Shklovskii1984}. The bulk resistance is governed by a critical resistance (distance) $R_c$ ($r_c$) which is the smallest resistance (or equivalently the smallest separation) that still allows for an infinitely large network of bonds.
This percolation length is set by the bonding criterion: $B_c = 4 \pi \int_0^{r_c}r^2 N dr$, where $N$ is the density of sites in the system and $B_c$ is a number that determines how many bonds each site in the percolating network must connect to on average; $B_c \approx 2.7$ in three dimensions \cite{Shklovskii1984}.
Energy disorder is negligible when the inter-site separation is large and temperatures are high; conductivity due to $r$-percolation has been observed in organic semiconductors in this regime \cite{Gill1974, Rubel2004}. These conditions are assumed throughout this Letter. For smaller inter-site separations or lower temperatures, energy disorder plays a pivotal role. 
Our model could, in principle, be generalized and solved numerically to treat such situations.
\begin{figure}[ptbh]
 \begin{centering}
        \includegraphics[scale = 0.32,trim = 80 160 80 200, angle = -0,clip]{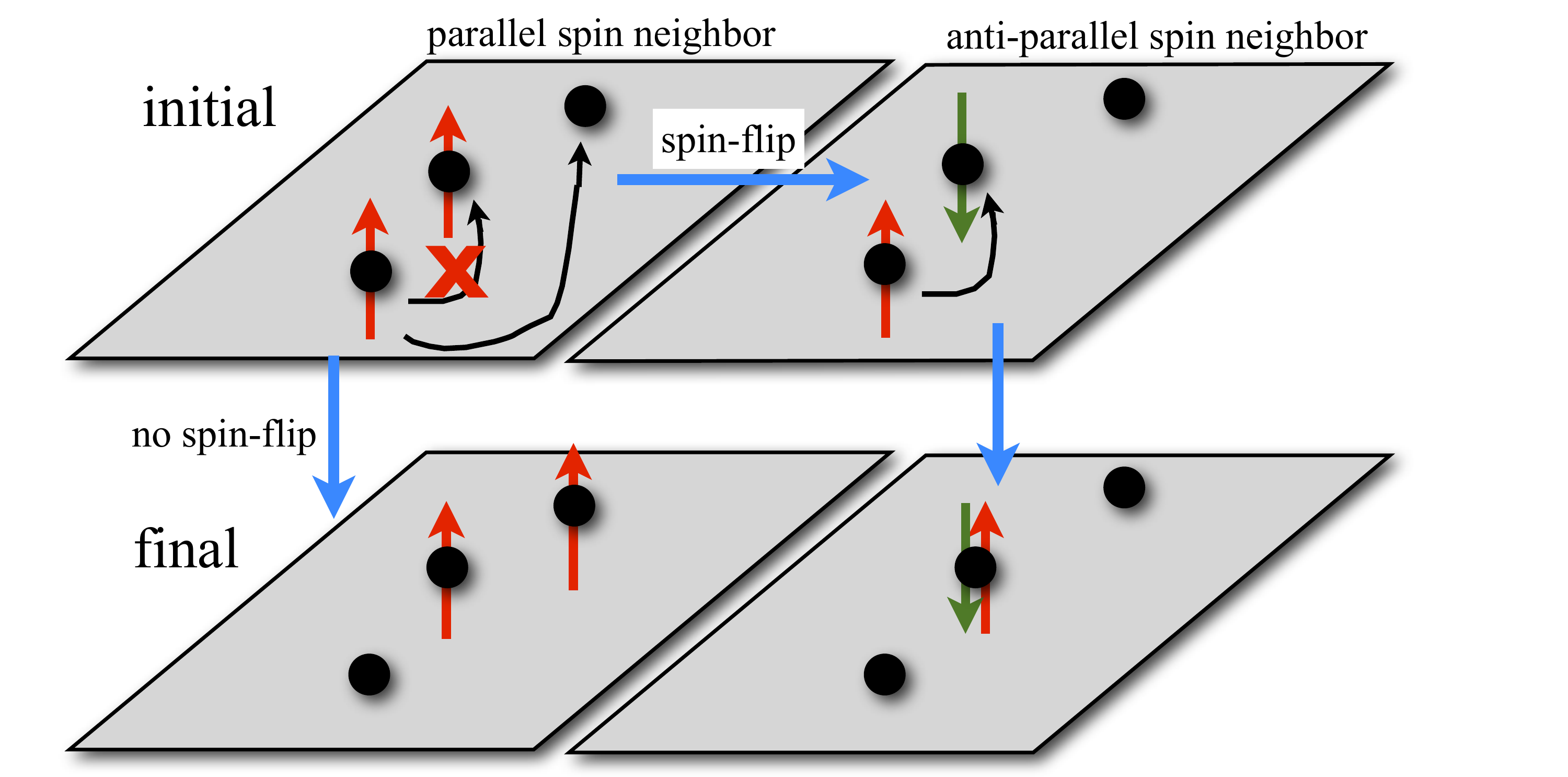}
        \caption[]
{Spin blocking in transport. 
Top: the initial situation for hopping when a carrier's nearest neighbor is occupied by a parallel spin (left) and anti-parallel spin (right).  Bottom: on the left  the spin-blocked carrier has made the more difficult further hop to an unoccupied site. On the right, the carrier successfully hops to the occupied nearest neighbor with anti-parallel spin. The carrier concentration is dilute enough such that if a carrier's nearest neighbor is occupied, it is improbable that its second nearest neighbor will be occupied.}\label{fig:cartoon}
        \end{centering}
\end{figure}

Spin affects electronic transport in hopping transport through the Pauli exclusion principle, as shown schematically in Fig~\ref{fig:cartoon}; double occupation of carriers on a single site is forbidden if their spins are parallel (P), but allowed if they are anti-parallel (AP) \cite{Bussac1993, Bobbert2007} (at the cost of a Coulomb interaction energy $U$). To clarify the role of spin blocking we use the simple picture of $U=0$, so a carrier with arbitrary spin is restricted from hopping to an occupied site with P-spin but may hop to a site occupied by an AP-spin just as it would to an unoccupied site.  The respective concentrations of these three types of sites are $N_{P}$, $N_{AP}$, and $N_{0}$. We consider carrier concentrations dilute enough to neglect hops to doubly occupied sites. Since carrier hopping to an occupied site with a P-spin is forbidden, the concentration of sites is effectively reduced to $N  - N_P$. In the absence of spin flips we would then write the bonding criterion as $B_c = 4 \pi \int_0^{r_c}r^2 N_{eff}' dr$ where $N_{eff}' = N  - N_P$.
The spin flip of a carrier at a site can be understood as a dynamical process that will cause the relative spin orientation between two singly occupied sites to change.  Hence, the hopping dynamics between two occupied sites is strongly dependent on spin flips.
If the total concentration of singly occupied sites is fixed at $N_s$, then at any given time the average densities of P-spin and AP-spin sites are $N_{s}/2$. Thus as a carrier attempts a hop to a singly occupied site, the probability of success will be 1/2, independent of spin effects. 
So, half the time the hop will be successful and  the density of sites for which these successful hops take place is $N_{AP} = N_s/2$. So as before the density of unrestricted hopping sites is $N_{eff}'$. 
We must now account for the situation that occurs the other one-half time in which the hopping attempt is foiled due to the singly occupied site being inhabited by a parallel spin, which occurs at $N_P = N_s/2$ sites. 

We introduce the possibility that the spin-blocked path can be opened by any process that alters the relative spin orientation between the two sites. The probability for the blockade to be lifted by the time the next hopping attempt takes place, $\tau_h$, is $p$. 
We thus modify the effective density of P sites  to be $[1-p] N_P$.
The resulting modification of the density of sites that can be hopped to, $N_{eff}$, within the model of Miller and Abrahams, accounts for the possibility of spin flips of spins located at singly occupied sites. 
Using the effective site density, we write the bonding criterion as
\begin{equation}\label{eq:bondingCriterion}
B_c =  \frac{4}{3} \pi a^3 y_c^3 (N - N_P) + 4 \pi a^3 N_P \int_0^{y_c}  y^2 p dy,
\end{equation}
where $y_c = r_c/a$ is the dimensionless critical length which dictates the threshold resistance $R_c = R_0 e^{2 y_c}$; $\tau_h =  v_0^{-1} e^{2 y}$ is the hopping time. A quantity $y_{c_0} = (3 B_c/4 \pi a^3 N)^{1/3}$ is defined as the critical inter-site spacing in the absence of all spin effects.
In general, $y_c$ cannot be isolated in Eq. (\ref{eq:bondingCriterion}) and the resultant MR can only be obtained  numerically, however, in the dilute carrier regime ($N_P \ll N$), the MR obeys the analytic expression
\begin{equation}\label{eq:generalMR}
\textrm{MR} \approx 2 \frac{1}{y^2_{c_1}} \frac{N_P}{N} \int_0^{y_{c_1}} y^2 [p(0)-p(H)]dy,
\end{equation}
where $y_{c_1} = y_{c_0} (1-N_P/N)^{-1/3}$ is the renormalized critical inter-site spacing.
The MR scales linearly with the fraction of singly occupied sites.

\begin{figure}[ptbh]
 \begin{centering}
        \includegraphics[scale = 0.54,trim = 5 0 0 0, angle = -0,clip]{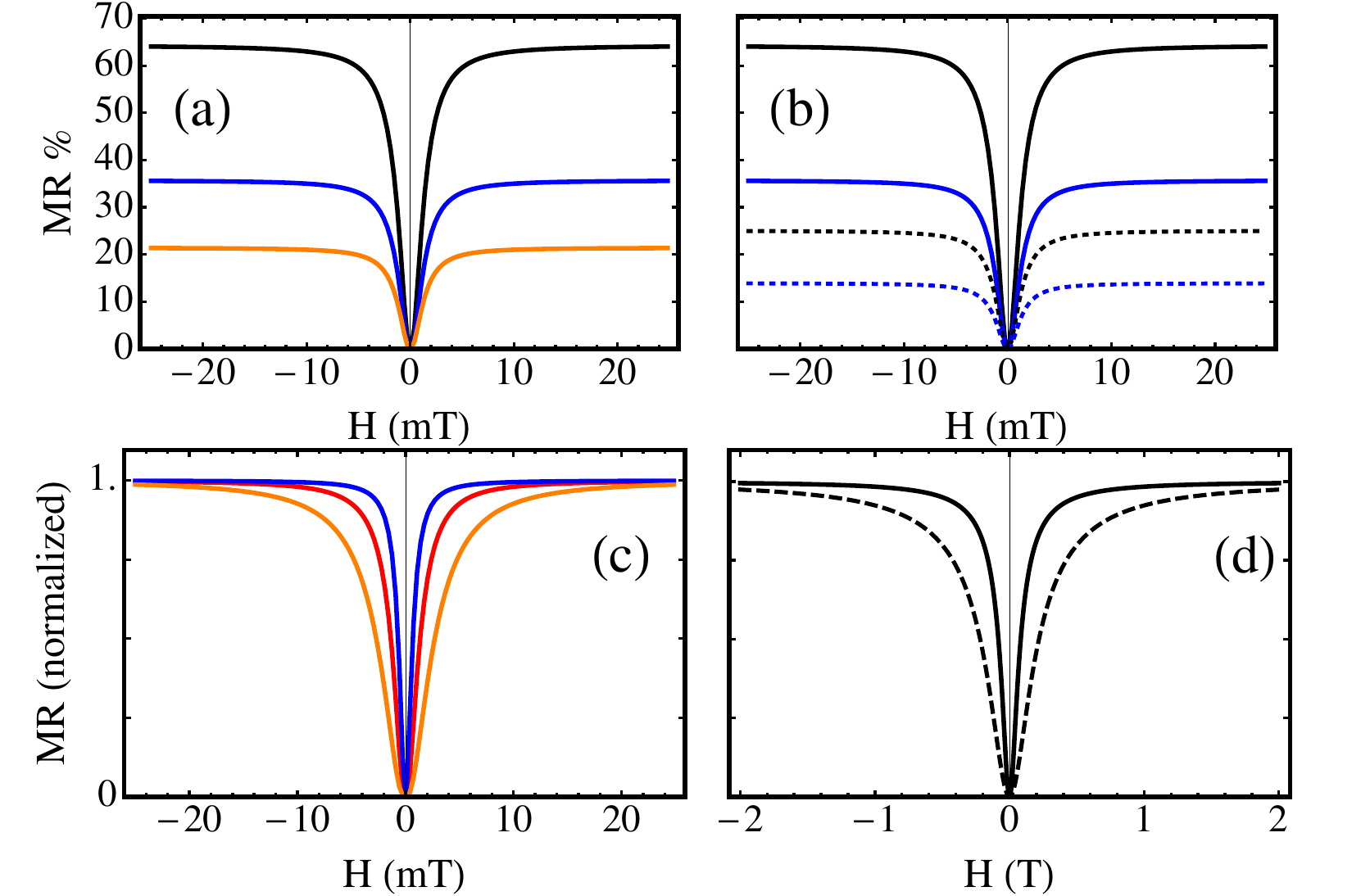}
        \caption[]
{(a) Calculated magnetoresistance \% for a polymer PPV derivative (orange line) with localization length $a = 3$ \AA ~ \cite{Martens2000}, the small molecule TNF (blue line) with localization length $a = 1.8$ \AA  ~\cite{Gill1974}, and a generic organic (black line) with localization length $a = 1$ \AA. Total site density is $N=10^{26}$ m$^{-3}$ and singly occupied site density is $N_P=10^{25}$ m$^{-3}$.
(b) Calculated magnetoresistance for generic organic (black) and TNF (blue) at site densities $N=10^{26}$ m$^{-3}$ (solid) and $N=2 \times 10^{26}$ m$^{-3}$ (dashed).
(c) Normalized magnetoresistance for generic organic at hyperfine field distribution widths ($h_I$) of 0.5 mT (blue line), 1 mT (red line), and 2 mT (orange line). Site densities are same as in (a). 
(d) Normalized magnetoresistance for generic organic at total site densities of $N=5 \times 10^{28}$ m$^{-3}$ (solid line) and $N=1 \times 10^{29}$ m$^{-3}$ (dotted line). Only hyperfine interaction is assumed; unless noted otherwise, all figures use $h_I = 1$ mT and $v_0 = 10^{12}$ s$^{-1}$.
}\label{fig:fig1}
        \end{centering}
\end{figure}

Most of the results reported below are based on the form of $p$ that is appropriate if the spin flips in these organic materials are caused by the hyperfine interaction (HI)\cite{Prigodin2006, Bobbert2007};  expressions and implications appropriate for spin flips caused by the spin-orbit interaction are summarized briefly at the end of this Letter.
Figure \ref{fig:fig1} emphasizes the main results of our theory.
Panel (a) shows our calculations of MR for three different organic semiconductors: the small molecule TNF which has electrons as carriers (blue), the polymer derivative of PPV (orange) with hole carriers, and a generic material (black) that possesses a smaller localization length (1 \AA) than the other two (1 \AA $<$ a$_{\textrm{TNF}} <$a$_{\textrm{PPV}}$ as given in figure caption).
This result suggests that organic materials with small localization lengths yield the largest MR.
Panel (b) contains our calculations for different site concentrations and indicates that dilute site concentrations are more magnetoresistive.
Panel (c) and (d) demonstrate the different MR structures and their dependence on the hyperfine field and site concentration obtainable from our theory. These results are derived and discussed in the remaining portion of this Letter.

\emph{The case of fast hopping} -
When hopping is faster than the hyperfine precession frequency, the carrier spin experiences a random field for the short duration of time that it resides at a site. The spin-flip rate is identical to the well known spin relaxation rate from HI in the motional narrowing regime \cite{Movaghar1977}:
\begin{equation}\label{eq:spinRelaxation}
\frac{1}{\tau_s} = \frac{v^2}{v_H^2 + \tau_h^{-2}}\frac{1}{\tau_h}.
\end{equation}
$v_H$ and $v$ are precession frequencies due to the external field $H$ and the in-plane internal hyperfine fields, of strength $h$.
The probability for the P-spin to flip at the next hop is $p(\tau_s) = 1 - e^{-\tau_h/\tau_s}$, which will permit a hop to the target site. 
This is a condition met in part when the density of sites is high. 
In this case the probability for a spin flip is $p(\tau_s) \approx \tau_h/\tau_s$ and the MR takes the form
\begin{equation}\label{eq:fastHopping}
\textrm{MR} \approx 2 \frac{1}{y^2_{c_1}} \frac{N_P}{N} \int_0^{y_{c_1}} y^2 \tau_h( \frac{1}{\tau_s(0)}-\frac{1}{\tau_s(H)}) dy.
\end{equation}
The hyperfine fields are random at each site so a correct description of the MR involves an average over the Gaussian distribution of these fields.

The resulting expression of Eq. (\ref{eq:fastHopping}) is cumbersome but can be simplified considerably by using the typical assumption that $y_{c_1} \gg 1$.
The resulting MR response, averaged over the Gaussian distribution of hyperfine fields with width $h_I$,  is
\begin{equation}\label{eq:osaka}
\langle \textrm{MR} \rangle =
\frac{1}{2}  \frac{N_{P}}{N} v_I^2 \tau^2_c\big[1 - \frac{1}{v_H^2 \tau^2_c}\ln(1+v_H^2 \tau^2_c) \big],
\end{equation}
where $\tau_c$ is the hopping time at the critical radius and $v_I$ is the precession frequency corresponding to the field $h_I$.
The positive MR can be understood by considering the field dependence of the relaxation mechanism; {\it e.g.} an increasing field \emph{suppresses} spin relaxation via HI which makes the spin blockade more effective. 
The dependence of MR on magnetic field here is identical to an earlier calculation for amorphous semiconductors performed in the fast hopping limit \cite{Osaka1979}. 
Such agreement suggests our model provides an accurate description of spin relaxation induced MR. 

\emph{The case of slow hopping} - The condition $1/v_I\gg\tau_h $ may not always be suitable for organic systems since the mobilities are so low. In positionally disordered systems the low mobility is entailed by low site concentrations. As the site concentration is reduced, the site separations increase and the carrier hopping rate is reduced, leading to the condition $1/v_I < \tau_c$.  
During the requisite waiting time to hop, the carrier spin at $i$ and target-site spin at $j$ experience the applied field and their respective hyperfine fields $h_i$ and $h_j$. 
Given two spins initially P aligned, the different hyperfine fields at the two sites rotate the spins to produce the possibility of AP alignment. 
We interpret this as a spin flip at either site. The time-averaged probability that the next hop is successful is $p(H) = p_{ij}+p_{ji}$ where $p_{ij}$ is the probability for the carrier at site $i$ to be opposite its initial state while the carrier at site $j$ remains in its initial state \cite{Sheng2006a, Shankar1994}:
\begin{equation}
p_{ij}= \frac{1}{2}\frac{h_i^2}{h_i^2+H^2} \Big[1 - \frac{1}{2}\frac{h_j^2}{h_j^2+H^2} )\Big].
\end{equation}
The second term, $p_{ji}$ is the reverse possibility.
The MR, from Eq. (\ref{eq:generalMR}), is ascertained to be
\begin{equation}\label{eq:slowHopping}
\textrm{MR} \approx \frac{1}{3} y_{c_1} \frac{N_P}{N} \frac{H^4}{(h_i^2+H^2)(h_j^2+H^2)}.
\end{equation}
We note that the MR is independent of the hopping rate, which contrasts starkly with the fast hopping case.
We now explore the MR lineshapes for these two cases.

\begin{figure}[ptbh]
\label{fig:gan}  \begin{centering}
        \includegraphics[scale = 0.62,trim = 0 0 0 00, angle = -0,clip]{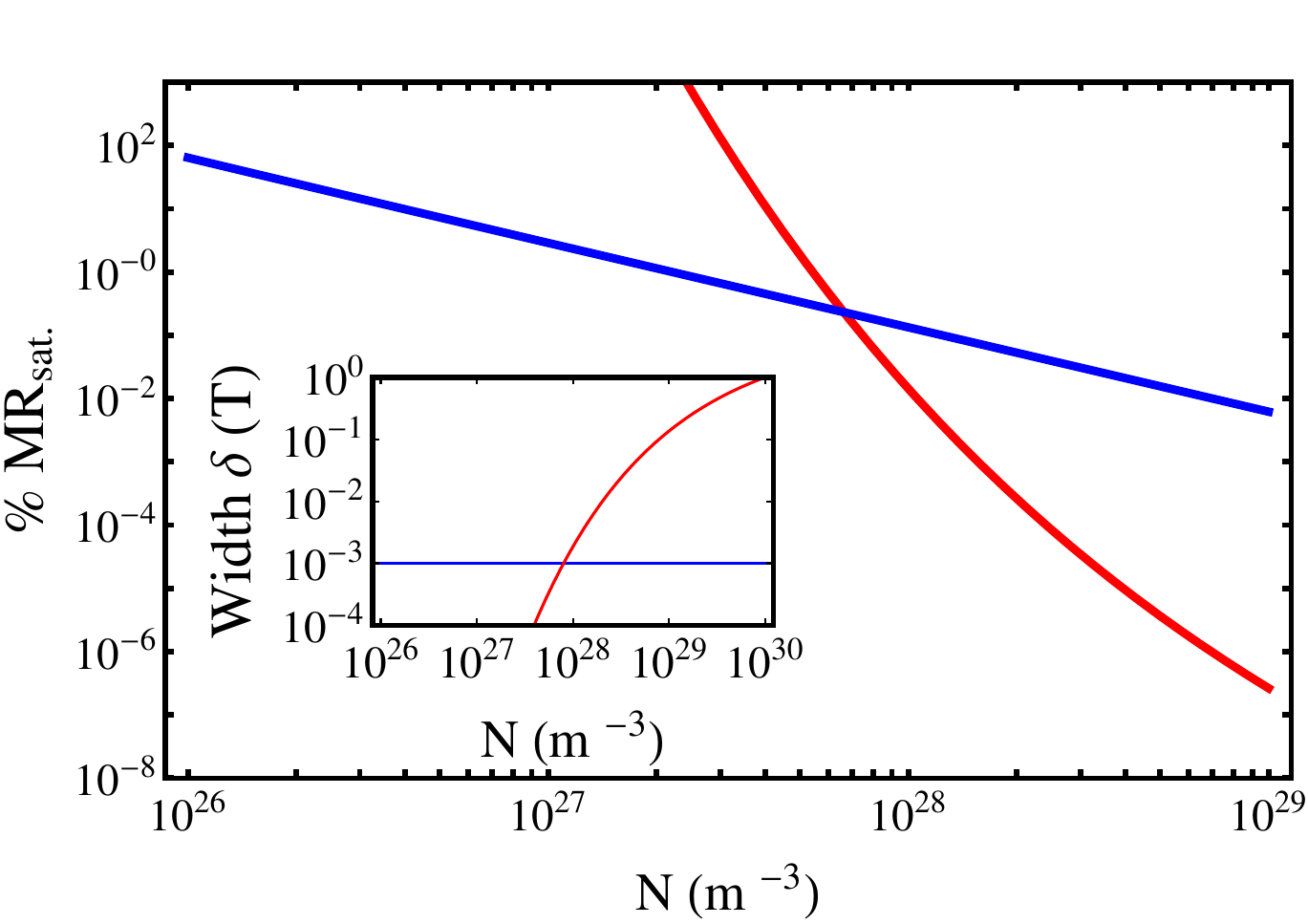}
        \caption[]
{Percentage saturated magnetoresistance versus site concentration. Blue line: slow hopping using Eq. (\ref{eq:mrSat}); red line: fast hopping using high field limit of Eq. (\ref{eq:osaka}) with $h_I = 1$ mT, $v_0 = 10^{12}$ s$^{-1}$, $a = 1$ \AA, and $N_P = 10^{25}$ m$^{-3}$.
Inset: Magnetoresistance width, $\delta$, as a function of site concentration for same parameters.}\label{fig:concentration}
        \end{centering}
\end{figure}

\emph{The magnetoresistance lineshape} - 
Two characteristic features to quantify MR are its value at high magnetic fields, MR$_{\textrm{sat.}}$, and its width, $\delta$, which we define as the half-width at half-maximum. The cross-over from slow to fast hopping can clearly be seen by viewing the saturated MR in Figure \ref{fig:concentration} and the MR width in the inset of the same figure. 
Slow hopping which results here from large inter-site distances, is conducive to large values of MR. 
The formula for $\textrm{MR}_{\textrm{sat.}}$ in the low site density regime, derived from Eq. (\ref{eq:slowHopping}), is
\begin{equation}\label{eq:mrSat}
\textrm{MR}_{\textrm{sat.}} \approx \frac{1}{3} \Bigg(\frac{3 B_c}{4 \pi a^3}\Bigg)^{1/3} \frac{N_P}{N^{4/3}}.
\end{equation}
Remarkably, the saturated MR is independent of HI.
We note that in the deuterated PPV experiments of Ref. \onlinecite{Nguyen2010}, $\delta$ of the magnetoluminesence decreased when the hyperfine field was reduced whereas the high field magnetoluminesence was nearly unchanged. This result is consistent with our theory. 
In the fast hopping limit, the site concentration dependence is even stronger since $\textrm{MR}_{\textrm{sat.}}$ contains an exponential dependence on $N$ through $ \tau_c$.
The MR widths also take on very different behaviors which are discerned from their MR expressions above and shown in the inset of Figure \ref{fig:concentration}. 
In the fast hopping region, the width is \emph{independent} of the hyperfine interaction but strongly dependent on the hopping rate.
Note that the width for fast hopping is larger - this is due to the quicker hopping rate which results in greater fields being required to suppress HI spin relaxation. 
MR widths as large as 40 mT - much greater than the hyperfine fields present - have been measured \cite{Bloom2007, Gomez2010}; Fig.~\ref{fig:concentration} suggests these large widths are related to the hopping rate and not the hyperfine field.
Often it has been assumed \cite{Wagemans2008, Rolfe2009, Nguyen2010} that $\delta$ must depend on $h_I$. We find this is only true in the slow hopping case where the MR curve roughly follows that of a Lorentzian of width $h_I$.

\emph{Spin-orbit coupling effects} - 
For fast hopping, Eq. (\ref{eq:fastHopping}) is quite general in that any spin relaxation mechanism can be included for $\tau_s$.
Here we consider the influence of spin-orbit coupling (SOC);
a recent calculation of SOC in several organic solids suggests that SOC may be significant \cite{Yu2011}.
Additionally SOC manifests itself by producing inhomogeneous g-factors (IG).
The result of SOC produces a spin relaxation rate: $\tau_s^{-1} = [b^2 v_H^2 (v_H^2 + \tau_h^{-2})^{-1}+\gamma^2]\tau_h^{-1}$
where $\gamma$ determines the SOC strength, $b = \sqrt{3/10} \delta g$, and $\delta g$ is known to be proportional to $\gamma$ \cite{Movaghar1977}. 
The field-independent portion reduces the total MR by $e^{-\gamma^2}$ but leads to no other qualitative change.
This result is expected since SOC is field-independent and has been observed in Alq$_3$ doped with Iridium \cite{Prigodin2006}. 
IG, which increases with increasing field, leads to negative MR:
 $-\frac{1}{2}  \frac{N_{P}}{N} b^2 \ln(1+v_H^2 \tau^2_c)$
where $(v_I + v_H)\tau_c \ll 1$. At low fields this effect is expected to be small compared to HI-induced MR.
Recently, IG was studied in the slow hopping regime \cite{Wang2008} but $\delta g$ was found to be unrealistically high to explain the magnetic field effects \cite{Wagemans2010}.

The prediction of large MR in the slow hopping regime necessitates measurements over a controlled and wide range of site concentrations.
We now discuss experimental strategies to observe our theoretical predictions.
For $r$-percolation theory to be valid, the organic system must possess a low density of molecular sites and this density must be controllable. 
Conduction via $r$-percolation was identified in TNF films by measuring the electron mobility through time-of-flight experiments \cite{Gill1974, Rubel2004}.
The molecular density of TNF could be carefully controlled by dispersing TNF in a polyester host that did not alter the transport properties.
We suggest similar experiments to look for the magnetoresistive dependences on hopping rate described in this Letter.
We conclude by noting that this theory has implications for MR effects in amorphous semiconductors \cite{Movaghar1977} and colloidal quantum dots \cite{GuyotSionnest2007},  as well as for spin diffusion in organic spin valves \cite{Bobbert2009}.

This work was supported by an ARO MURI. We acknowledge stimulating discussions with M. Wohlgenannt.

\end{document}